\documentclass{appolb}
\usepackage{graphicx}
\begin{document}
\title{Relating eccentricity fluctuations to density fluctuations in heavy-ion collisions\thanks{Presented at XXV Cracow Epiphany Conference on Advances in Heavy Ion Physics.}}
\author{Rajeev S. Bhalerao$^a$, Giuliano Giacalone$^b$, Pablo Guerrero-Rodr\'iguez$^c$, Matthew Luzum$^d$, Cyrille Marquet$^e$, Jean-Yves Ollitrault$^b$
  \address{$^a$Department of Physics, Indian Institute of Science Education and Research (IISER), Homi Bhabha Road, Pune 411008, India\\
    $^b$Institut de physique th\'eorique, Universit\'e Paris Saclay, CNRS, CEA, F-91191 Gif-sur-Yvette, France\\
    $^c$CAFPE and Departamento de F\'isica Te\'orica y del Cosmos, Universidad de Granada, E-18071 Campus de Fuentenueva, Granada, Spain\\
    $^d$Instituto de F\'{i}sica, Universidade de S\~{a}o Paulo, C.P.
  66318, 05315-970 S\~{a}o Paulo, Brazil\\
    $^e$CPHT, \'Ecole Polytechnique, CNRS, Route de Saclay, 91128 Palaiseau, France}
}
\maketitle
\begin{abstract}
  The magnitude of anisotropic flow in a nucleus-nucleus collision is determined by the energy density field, $\rho(x,y,z)$, created right after the collision occurs.
  Specifically, elliptic flow, $v_2$, and triangular flow, $v_3$, are proportional to the anisotropy coefficients $\varepsilon_2$ and $\varepsilon_3$, which are functionals of $\rho$. 
  We express the mean and the variance of $\varepsilon_2$ and $\varepsilon_3$ as a function of the 1- and 2-point functions of $\rho$. 
  These results generalize results obtained previously, that were valid only for central collisions, or only for identical point-like sources.
We apply them to the color glass condensate effective theory, using the recently derived expression of the 2-point function. 
\end{abstract}
  
\section{Introduction}
Anisotropic flow is central to the phenomenology of heavy-ion collisions. It is the phenomenon that converts the anisotropy of the initial energy-density profile created in the collision into final-state momentum anisotropy, which is accurately measured by the detectors~\cite{Adams:2004bi,Adler:2003kt,Aamodt:2010pa,ATLAS:2012at,Chatrchyan:2012ta}.
In the final state, anisotropy is characterized by the Fourier coefficients of the azimuthal distribution $P(\varphi_p)$ of outgoing particles~\cite{Luzum:2011mm}:
\begin{equation}
  \label{defvn}
v_n=\int_{0}^{2\pi}e^{in\varphi_p}P(\varphi_p)d\varphi_p.
\end{equation}
In the initial state, anisotropy is characterized by the Fourier coefficients~\cite{Teaney:2010vd,Qiu:2011iv} $\varepsilon_n$ of the initial energy density profile at mid-rapidity $\rho({\bf s})$, where ${\bf s}$ labels a point in the transverse plane:\footnote{We shall not be dealing with the longitudinal dynamics of the system, which is not important for our discussion.}
\begin{equation}
  \label{defepsn1}
  \varepsilon_n=\frac{\int_{\bf s}e^{in\varphi}|{\bf s}|^n \rho({\bf s})}
  {\int_{\bf s}|{\bf s}|^n \rho({\bf s})}.
\end{equation}
In this equation, and throughout this article, we use the short hand $\int_{{\bf s}}=\int {\rm d}x{\rm d}y$ for the integration over the transverse  plane.
$|{\bf s}|$ and $\varphi$ are polar coordinates of ${\bf s}$ in a centered coordinate system, to be defined below. 

Hydrodynamic simulations show that for the largest harmonics, $n=2$ (elliptic flow) and $n=3$ (triangular flow), $v_n$ is to a good approximation~\cite{Gardim:2011xv,Niemi:2012aj,Gardim:2014tya} proportional to $\varepsilon_n$, that is, $v_n = \kappa_n \varepsilon_n$, where $\kappa_n$ is a hydrodynamic response coefficient which depends mildly on the impact parameter of the collision at a given energy.
Therefore, the initial anisotropy $\varepsilon_n$ acts as the seed of anisotropic flow, $v_n$.
The two main effects producing a non-zero $\varepsilon_n$ are:
\begin{itemize}
\item{The almond shape of the overlap area between two nuclei for noncentral collisions, that generates a large $\varepsilon_2$~\cite{Ollitrault:1992bk}.}
\item{Event-to-event density fluctuations~\cite{Miller:2003kd}, that generate a non-zero $\varepsilon_2$ even in central collisions~\cite{Alver:2006wh}, and a non-zero $\varepsilon_3$ for all centralities~\cite{Alver:2010gr}.}
\end{itemize}
The measured $v_n$ is typically a rms average over many events at a fixed impact parameter $b$.
Therefore, the relevant quantity for phenomenology is the rms average of $\varepsilon_n$.

In this article, we express the rms averages of $\varepsilon_2$ and $\varepsilon_3$ in terms of the 1- and 2-point functions of the energy density field $\rho({\bf s})$, thus providing a direct link between models of initial conditions and quantities of phenomenological relevance. 
We carry out a statistical average over events, where the positions of the centers of the nuclei are fixed (which implies that the impact parameter $b$ is also fixed). 
The 1-point function is the average of $\rho({\bf s})$, which we denote by $\langle\rho({\bf s})\rangle$.
The 2-point function $S({\bf s}_1,{\bf s}_2)$ characterizes the variance of the fluctuations. Decomposing the density field as $\rho({\bf s})=\langle\rho({\bf s})\rangle+\delta\rho({\bf s})$, where $\delta\rho({\bf s})$ is an event-by-event fluctuation, $S({\bf s}_1,{\bf s}_2)$ is defined  by
\begin{equation}
 \label{defS}
 S({\bf s}_1,{\bf s}_2)  \equiv \langle\delta\rho({\bf s}_1)\delta\rho({\bf s}_2)\rangle
 =\langle\rho({\bf s}_1)\rho({\bf s}_2)\rangle-\langle\rho({\bf s}_1)\rangle\langle\rho({\bf s}_2)\rangle.
\end{equation}
In~\cite{Blaizot:2014nia}, the rms values of $\varepsilon_n$ were obtained in terms of $\langle\rho({\bf s})\rangle$ and  $S({\bf s}_1,{\bf s}_2)$ in the specific case of central collisions with $b=0$.
We generalize these results to the general case $b\not=0$.
Results for $b\not=0$ have already been obtained in the specific case of identical point-like sources~\cite{Bhalerao:2006tp,Alver:2008zza,Bhalerao:2011bp}, corresponding to the following 2-point function~\cite{Gronqvist:2016hym}:
\begin{equation}
\label{Ssource}
S({\bf s}_1,{\bf s}_2)=C\langle\rho({\bf s}_1)\rangle \delta({\bf s}_1-{\bf s}_2),
\end{equation}
where $C$ is a constant proportionality factor. We generalize these results to an arbitrary $S({\bf s}_1,{\bf s}_2)$.
 We finally carry out a numerical implementation of our results in the case of the color glass condensate (CGC), whose 1- and 2-point functions were evaluated by Albacete {\it et al.\/}~\cite{Albacete:2018bbv}.

\section{Perturbative expansion of initial anisotropies}
\label{s:pert}
We expand $\varepsilon_2$ and $\varepsilon_3$  in powers of the density fluctuation $\delta\rho$. 
We first introduce the following shorthand notations, for any function $f({\bf s})$~\cite{Gronqvist:2016hym}:
\begin{eqnarray}
\label{notation}
\delta f &\equiv&\frac{1}{\langle E\rangle}\int_{\bf s} f({\bf s})\delta\rho({\bf s})\cr
\langle f\rangle &\equiv&\frac{1}{\langle E\rangle}\int_{\bf s} f({\bf s})\langle\rho({\bf s})\rangle,
\end{eqnarray}
where $\langle E\rangle$ is defined by:
\begin{equation}
\label{defeav}
\langle E\rangle=\int_{\bf s} \langle\rho({\bf s})\rangle.
\end{equation}
Thus $\langle f\rangle$ is the average value of $f({\bf s})$ with the weight $\langle\rho({\bf s})\rangle$.
Throughout this article, we refer to $E$ as to the ``total energy'', but it represents the density of energy per unit longitudinal length. 

We work in a coordinate frame where the origin of the transverse plane lies at the center of the average energy density, so that
\begin{equation}
\label{deforigin} 
\langle {\bf s}\rangle=0.
\end{equation}
Due to fluctuations, the center of the distribution $\rho({\bf s})$, which we denote by ${\bf s}_0$, fluctuates event to event.
It is defined by
\begin{equation}
  {\bf s}_0\equiv\frac{\int_{\bf s}{\bf s}\rho({\bf s})}{\int_{\bf s}\rho({\bf s})}\simeq
  \frac{\int_{\bf s}{\bf s}\delta\rho({\bf s})}{\int_{\bf s}\langle\rho({\bf s})\rangle}=\delta{\bf s},
  \end{equation}
    where we have neglected $\delta\rho$ in the denominator (which amounts to expanding ${\bf s}_0$ to leading order in $\delta\rho$), and used the notation of Eq.~(\ref{notation}) in the last equality.
    
Throughout this article, we use the complex notation ${\bf s}=x+iy=|{\bf s}|e^{i\varphi}$.
This allows us to rewrite Eq.~(\ref{defepsn1}) as:
\begin{equation}
  \label{defepsn2}
  \varepsilon_n=\frac{\int_{\bf s}({\bf s}-\delta{\bf s})^n \rho({\bf s})}
  {\int_{\bf s}|{\bf s}-\delta{\bf s}|^n \rho({\bf s})}, 
\end{equation}
where the recentering correction $\delta{\bf s}$ ensures that anisotropies are evaluated in a centered frame~\cite{Teaney:2010vd,Alver:2006wh}.
Our goal is to evaluate the mean value and the variance of $\varepsilon_n$ to order 2 (lowest non-trivial order) in the fluctuations. 

For $n=2$, Eq.~(\ref{defepsn2}) can be rewritten as:
\begin{equation}
\label{flucteps2}
  \varepsilon_2=\frac{\langle {\bf s}^2\rangle+\delta {\bf s}^2-(\delta {\bf s})^2}
  {\langle {\bf ss^*}\rangle+\delta {\bf ss^*}-(\delta {\bf s})(\delta {\bf s^*})},
\end{equation}
where ${\bf s^*}$ is the complex conjugate of ${\bf s}$, and we have used the notation introduced in Eq.~(\ref{notation}).
The terms in $\delta$ are proportional to the density fluctuation. 
If one neglects them, $\varepsilon_2$ reduces to the eccentricity of the mean density profile, which we denote by $\bar\varepsilon_2$:
\begin{equation}
  \label{bareps2}
\bar\varepsilon_2\equiv\frac{\langle {\bf s}^2\rangle}{\langle {\bf ss^*}\rangle}=\frac{\langle {\bf s}^2\rangle}{\langle |{\bf s}|^2\rangle}.
\end{equation}
Expanding Eq.~(\ref{flucteps2}) and keeping all terms up to order 2, we obtain:
\begin{equation}
  \label{perteps2}
  \varepsilon_2=\bar\varepsilon_2+\frac{\delta {\bf s}^2}{\langle |{\bf s}|^2\rangle}
-\bar\varepsilon_2\frac{\delta {\bf ss^*}}{\langle |{\bf s}|^2\rangle}
  -\frac{(\delta {\bf ss^*})(\delta {\bf s}^2)}{\langle |{\bf s}|^2\rangle^2}
    +\bar\varepsilon_2\frac{(\delta {\bf ss^*})^2}{\langle |{\bf s}|^2\rangle^2}
  -\frac{(\delta {\bf s})^2}{\langle |{\bf s}|^2\rangle}
  +\bar\varepsilon_2\frac{(\delta {\bf s})(\delta {\bf s}^*)}{\langle |{\bf s}|^2\rangle}.
\end{equation}
One easily checks that this equation satisfies rotational symmetry: Both sides are multiplied by $e^{2i\alpha}$ under the transformation ${\bf s}\to{\bf s}e^{i\alpha}$. 
The right-hand side of Eq.~(\ref{perteps2}) contains two terms of order 1 in the fluctuations, and four terms of order 2.

The mean anisotropy $\langle\varepsilon_2\rangle$ is obtained by averaging Eq.~(\ref{perteps2}) over events.\footnote{Note that we use the same angular brackets to denote an average over events, or an average value taken with the mean density profile, as in Eq.~(\ref{notation}). There should be no confusion depending on the context.}
One-point averages of the type $\langle \delta f\rangle$ vanish by definition of $\delta\rho$.
Therefore, only the terms of order 2 contribute:
\begin{equation}
  \label{pertbareps2}
  \langle\varepsilon_2\rangle=\bar\varepsilon_2
    -\frac{\langle(\delta {\bf ss^*})(\delta {\bf s}^2)\rangle}{\langle |{\bf s}|^2\rangle^2}
    +\bar\varepsilon_2\frac{\langle(\delta {\bf ss^*})^2\rangle}{\langle |{\bf s}|^2\rangle^2}
  -\frac{\langle(\delta {\bf s})^2\rangle}{\langle |{\bf s}|^2\rangle}
  +\bar\varepsilon_2\frac{\langle(\delta {\bf s})(\delta {\bf s}^*)\rangle}{\langle |{\bf s}|^2\rangle}.
\end{equation}
Thus, the mean anisotropy $\langle\varepsilon_2\rangle$ differs from the anisotropy of the mean density $\bar\varepsilon_2$ by terms of order 2 in the fluctuations.
Note that the last two terms come from the recentering correction.
The numerators involve 2-point averages of the type $\langle \delta f\delta g\rangle$, where $f$ and $g$ are two functions of ${\bf s}$.
Such averages can be readily expressed in terms of the 2-point function using Eqs.~(\ref{defS}) and (\ref{notation}):
\begin{equation}
\label{2pointav}
\langle\delta f\delta g\rangle=
\frac{1}{\langle E\rangle^2}\int_{{\bf s}_1,{\bf s}_2}
f({\bf s}_1)g({\bf s}_2)S({\bf s}_1,{\bf s}_2).
\end{equation}

We now evaluate the variance of $\varepsilon_2$ fluctuations, which we define by~\cite{Giacalone:2019kgg}:
\begin{equation}
  \label{defsigma2}
  \sigma^2\equiv\left\langle\left|\varepsilon_2-\langle\varepsilon_2\rangle\right|^2\right\rangle
=
\left\langle\varepsilon_2\varepsilon_2^*\right\rangle-
\left\langle\varepsilon_2\right\rangle\left\langle\varepsilon_2^*\right\rangle. 
\end{equation}
Only the terms of order 1 in Eq.~(\ref{perteps2}) contribute to $\sigma^2$.
The reason is that the terms of order 2 give the same contribution to
$\left\langle\varepsilon_2\varepsilon_2^*\right\rangle$ and
$\left\langle\varepsilon_2\right\rangle\left\langle\varepsilon_2^*\right\rangle$, which cancels out in the difference. 
One thus obtains
\begin{equation}
\label{pertsigma2}
\sigma^2=
\frac{\left\langle\delta {\bf s}^2\delta {\bf s}^{{\bf *}2}\right\rangle
  +|\bar\varepsilon_2|^2\left\langle(\delta {\bf ss^*})^2\right\rangle
  -2\,{\rm Re}\left[\bar\varepsilon_2\left\langle\delta {\bf ss^*}\delta {\bf s}^{{\bf *}2}\right\rangle\right]
}{\langle |{\bf s}|^2\rangle^2},
\end{equation}
where ${\rm Re}[f]\equiv (f+f^*)/2$ denotes the real part of $f$. 
The rms value of $\varepsilon_2$, usually denoted by $\varepsilon_2\{2\}$, is then given by:
\begin{equation}
  \label{eps22}
  \varepsilon_2\{2\}^2\equiv\left\langle\varepsilon_2\varepsilon_2^*\right\rangle=|\langle\varepsilon_2\rangle|^2+\sigma^2.
\end{equation}

We discuss now the triangularity, $\varepsilon_3$.
We restrict our study to symmetric collisions, for which $\phi\to\phi+\pi$ symmetry implies $\langle\varepsilon_3\rangle=0$ for all centralities.
In order to obtain the variance of $\varepsilon_3$ fluctuations to order 2 in the fluctuations, it suffices to keep only terms of order 1 in the definition of $\varepsilon_3$.
Using Eqs.~(\ref{notation}) and (\ref{defepsn2}), one obtains:
\begin{equation}
\label{flucteps3}
\varepsilon_3=\frac{\delta{\bf s}^3-
  3\langle {\bf s}^2\rangle\delta {\bf s}}
  {\langle |{\bf s}|^{3}\rangle}.
\end{equation}
The variance is obtained by multiplying with $\varepsilon_3^*$ and averaging over events:
\begin{equation}
\label{pertsigma3}
\varepsilon_3\{2\}^2\equiv 
\langle\varepsilon_3\varepsilon_3^*\rangle
=\frac{\left\langle\delta {\bf s}^3\delta {\bf s}^{{\bf *}3}\right\rangle
+9\left|\langle {\bf s}^2\rangle\right|^2
\left\langle\delta {\bf s}\delta{\bf s^*}\right\rangle
-6\,{\rm Re}\left[\langle {\bf s}^{2}\rangle\left\langle\delta {\bf s}\delta {\bf s}^{{\bf *}3}\right\rangle\right]
}{\langle |{\bf s}|^{3}\rangle^2}.
\end{equation}
Equations~(\ref{pertbareps2}), (\ref{pertsigma2}) and  (\ref{pertsigma3}), together with Eq.~(\ref{2pointav}), express the mean and the variance of $\varepsilon_2$ and $\varepsilon_3$ in terms of the 1- and 2-point functions of the density field $\rho({\bf s})$, to leading order in the fluctuations.
This is our first important result. 

Approximate expressions were previously used in~\cite{Giacalone:2019kgg}, where only the first term in the right-hand sides of Eq.~(\ref{pertbareps2}), (\ref{pertsigma2}) and  (\ref{pertsigma3}) was kept. 
The full expressions derived here are more accurate for non-central collisions. 
Let us briefly discuss the origin of the additional terms. 
If one replaces $\langle {\bf s}^2\rangle=\bar\varepsilon_2\langle |{\bf s}|^2\rangle$ in Eq.~(\ref{pertsigma3}), this equation shows some similarity with Eq.~(\ref{pertsigma2}), in the sense that in both equations, the second term is positive and proportional to $|\bar\varepsilon_2|^2$ while the last term is negative and proportional to $\bar\varepsilon_2$.
In Eq.~(\ref{pertsigma2}), these terms originate from the fluctuations in the system size (denominator of Eq.~(\ref{flucteps2})), while in Eq.~(\ref{pertsigma3}), they originate from the recentering correction. 

\section{Short-range correlations}
\label{s:short}

All the above results involve 2-point averages of the type (\ref{2pointav}).
We now explain how these averages are evaluated in practice.
We first change variables to ${\bf s}_1={\bf s}+{\bf r}/2$ and ${\bf s}_2={\bf s}-{\bf r}/2$:
\begin{equation}
\label{2pointav2}
\langle\delta f\delta g\rangle=
\frac{1}{\langle E\rangle^2}\int_{{\bf s},{\bf r}}
f\left({\bf s}+\frac{\bf r}{2}\right)g\left({\bf s}-\frac{\bf r}{2}\right)S\left({\bf s}+\frac{\bf r}{2},{\bf s}-\frac{\bf r}{2}\right).
\end{equation}
We assume short-range correlations, so that only values of ${\bf r}$ much smaller than the nuclear radius contribute in this integral. 
Then, if $f$ and $g$ are slowly-varying functions, 
one can make the approximations $f({\bf s}+{\bf r}/2)\simeq f({\bf s})$ and $g({\bf s}-{\bf r}/2)\simeq g({\bf s})$ in Eq.~(\ref{2pointav2}) and integrate over ${\bf r}$.
We introduce the notation~\cite{Giacalone:2019kgg}
\begin{equation}
\label{eq:2p}
  \xi({\bf s})\equiv
  \int_{\bf r}S\left( {\bf s}+\frac{{\bf r}}{2},{\bf s}-\frac{{\bf r}}{2}\right).
\end{equation}
The integral of $\xi$ is the variance of $E$:
\begin{equation}
\label{defevar}
\int_{\bf s} \xi({\bf s})=\int_{{\bf s}_1,{\bf s}_2}S({\bf s}_1,{\bf s}_2)=\Delta E^2.
\end{equation}
Thus the function $\xi({\bf s})$ represents the ``density of variance'', in the same way as $\langle\rho({\bf s})\rangle$ represents the density of mean energy.

We denote the average value of $f({\bf s})$ with the weight $\xi({\bf s})$ by $\{f\}$, to distinguish it from $\langle f\rangle$, which is an average with the weight $\langle\rho({\bf s})\rangle$:
\begin{equation}
\label{notation2}
\{ f\} \equiv\frac{\int_{\bf s} f({\bf s})\xi({\bf s})}{\int_{\bf s}\xi({\bf s})}.
\end{equation}
With these notations, the 2-point average (\ref{2pointav2}) becomes:
\begin{equation}
  \label{2pointav3}
  \langle\delta f\delta g\rangle=\frac{\Delta E^2}{\langle E\rangle^2}
 \{ fg\}.
\end{equation}
We thus rewrite our results (\ref{pertbareps2}), (\ref{pertsigma2}) and (\ref{pertsigma3}) as:
\begin{eqnarray}
  \label{results1}
  \langle\varepsilon_2\rangle&=&\bar\varepsilon_2
  +\frac{\Delta E^2\left(
  -\{  {\bf s}^3{\bf s^*}\}
    +\bar\varepsilon_2\{ {\bf s}^2{\bf s}^{{\bf *}2}\}
    -\{ {\bf s}^2\}\langle {\bf ss^*}\rangle
    +\langle {\bf s}^2\rangle \{ {\bf ss^*}\}
    \right)}
  {\langle E\rangle^2\langle |{\bf s}|^2\rangle^2}\cr
  \sigma^2&=&
  \frac{\Delta E^2\left((1+|\bar\varepsilon_2|^2)\{ {\bf s}^2{\bf s}^{{\bf *}2}\}
    -2\, {\rm Re}\left[\bar\varepsilon_2\{ {\bf s s}^{{\bf *}3}\}\right]
    \right)}
  {\langle E\rangle^2\langle |{\bf s}|^2\rangle^2}\cr
  \varepsilon_3\{2\}^2&=&
  \frac{\Delta E^2\left(\{{\bf s}^3{\bf s}^{{\bf *}3}\}
+9\left|\langle {\bf s}^2\rangle\right|^2\{ {\bf ss^*}\}
-6\,{\rm Re}\left[\langle {\bf s}^{2}\rangle\{ {\bf s}{\bf s}^{{\bf *}3}\}\right]
\right) }
       {\langle E\rangle^2\langle |{\bf s}|^3\rangle^2}.
\end{eqnarray}
This is our main result, which expresses the mean and the variance of $\varepsilon_2$ and $\varepsilon_3$ as a function of the mean $\langle\rho({\bf s})\rangle$ and the variance $\xi({\bf s})$ of the density field. 
The first term in the right-hand side of these equations is in general the dominant term, and the other terms are subleading terms which are all of the same order of magnitude.
Their relative magnitudes can be easily evaluated if $\xi$ and $\langle\rho\rangle$ are identical, Gaussian profiles.
Then, Wick's theorem implies $\{ {\bf s}^2{\bf s}^{{\bf *}2}\}=2 \{{\bf ss^*}\}^2+\{ {\bf s}^2\}\{{\bf s}^{{\bf *}2}\}\simeq 2 \{{\bf ss^*}\}^2$ and $\{  {\bf s}^3{\bf s^*}\}=3\{  {\bf s}^2\}\{  {\bf ss^*}\}$.
Therefore, the subleading terms are in the ratio $-3$, $2$, $-1$, $1$ for $\langle\varepsilon_2\rangle$, $1$ and $-3$ for $\sigma^2$, $1$ and $-2$ for $\varepsilon_3\{2\}^2$.
This result in a net negative correction to the leading term for all three quantities. 

We now check that Eqs.~(\ref{results1}) are compatible with results previously obtained, in two specific cases. 
The first case is that of central collisions, $b=0$.
Rotational symmetry then implies that all the terms in the right-hand side of these equations vanish, except for the first contribution to $\sigma^2$, and the first contribution to $\varepsilon_3\{2\}^2$.
The resulting expressions were previously obtained in~\cite{Blaizot:2014nia}.

The second case is that of independent, point-like sources carrying unit energy~\cite{Bhalerao:2006tp}.
If the number of sources follows a Poisson distribution, then, the 2-point function is given by Eq.~(\ref{Ssource}) which, together with Eq.~(\ref{eq:2p}), implies $\xi({\bf s})=C\langle\rho({\bf s})\rangle$. 
This in turn implies that $\{ f\}=\langle f\rangle$ for any function $f({\bf s})$.
The mean energy $\langle E\rangle$ and the variance $\Delta E^2$ are equal to the mean number of sources $\langle N\rangle$, and Eq.~(\ref{results1}) reduces to 
\begin{eqnarray}
  \label{results2}
  \langle\varepsilon_2\rangle&=&\bar\varepsilon_2
+  \frac{ 
  \bar\varepsilon_2\langle {\bf s}^2{\bf s}^{{\bf *}2}
  \rangle -\langle  {\bf s}^3{\bf s^*}\rangle 
}
  {  \langle N\rangle\langle |{\bf s}|^2\rangle^2}\cr
  \sigma^2&=&
\frac{
(1+|\bar\varepsilon_2|^2)\langle {\bf s}^2{\bf s}^{{\bf *}2}\rangle
    -2\,{\rm Re}\left[\bar\varepsilon_2\langle {\bf s s}^{{\bf *}3}\rangle\right]
    }
  {  \langle N\rangle\langle |{\bf s}|^2\rangle^2}\cr
  \varepsilon_3\{2\}^2&=&
\frac{
\langle{\bf s}^3{\bf s}^{{\bf *}3}\rangle
+9\left|\langle {\bf s}^2\rangle\right|^2\langle {\bf ss^*}\rangle
-6\,{\rm Re}\left[\langle {\bf s}^{2}\rangle\langle {\bf s}{\bf s}^{{\bf *}3}\rangle\right]
}
       {  \langle N\rangle\langle |{\bf s}|^3\rangle^2}.
\end{eqnarray}
Note that the recentering correction to $\langle\varepsilon_2\rangle$ (last two terms in the first line of Eq.~(\ref{results1})) vanishes in this case. 
The first two lines of Eq.~(\ref{results2}) agree with the result derived in~\cite{Bhalerao:2006tp}, while the third line corresponds to the result in~\cite{Bhalerao:2011bp}.
In these papers, however, the number of sources $N$ was assumed to be constant, as opposed to following a Poisson distribution. 
In other terms, the total energy was fixed.
The modifications of our results when the total energy is fixed will be derived below in Sec.~\ref{s:long}.

\section{Application to CGC effective theory}
\label{s:CGC}

In the CGC effective theory~\cite{Albacete:2018bbv}, the range of energy correlations induced by the QCD dynamics typically does not extend beyond a confinement scale of order $1$~fm, much smaller than the nuclear radius, so that the results of Sec.~\ref{s:short} apply.
The average density $\langle\rho({\bf s})\rangle$ and the density of variance $\xi({\bf s})$ depend on the saturation scales $Q_{A}({\bf s})$ and $Q_B({\bf s})$ of the two incoming nuclei~\cite{Giacalone:2019kgg}:
\begin{eqnarray}
\langle\rho({\bf s})\rangle &\!=&\! \frac{4}{3g^2} Q_{A}^2({\bf s}) Q_{B}^2({\bf s})\cr
  \xi({\bf s})
  &\!=&\!\frac{16\pi}{9 g^4}  Q_{A}^2({\bf s}) Q_{B}^2({\bf s})
 \!\left[\!Q_{A}^2({\bf s})\ln\!\left(\frac{Q_{B}^2({\bf s})}{m^2}\right)
  \!+Q_{B}^2({\bf s})\ln\!\left(\frac{Q_{A}^2({\bf s})}{m^2}\right)\right],
\end{eqnarray}
where $g$ is the coupling constant, and $m$ is an infrared cutoff which we take equal to the pion mass, $m=0.14$~GeV.
$Q_A^2({\bf s})$ and $Q_B^2({\bf s})$ are proportional to the thickness functions of the nuclei $T_A({\bf s})$ and $T_B({\bf s})$, which are obtained by integrating the nuclear density over the longitudinal coordinate~\cite{Miller:2007ri}.
The only free parameter in this approach is the proportionality coefficient or, equivalently, the saturation scale at the center of the nucleus, which we denote by $Q_{s0}$. 
In the numerical evaluation, we replace $\ln x$ with $\ln(1+x)$ in the above expression to ensure that the density of variance, $\xi$, is everywhere positive.
This is however inessential and the resulting modification is minor. 

\begin{figure*}[t]
\centerline{\includegraphics[width=12.5cm]{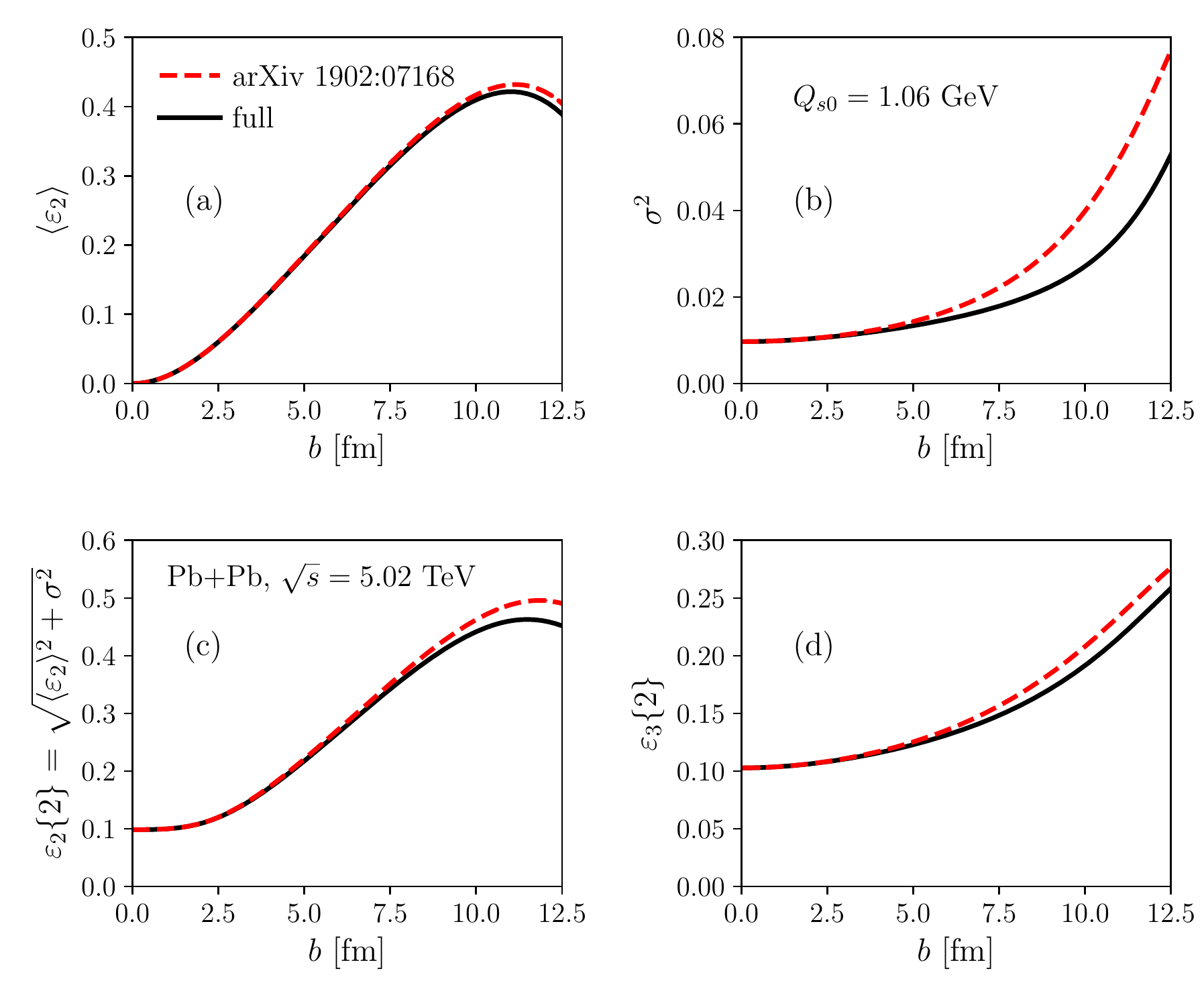}}
\caption{(a) $\langle\varepsilon_2\rangle$; (b) $\sigma^2$; (c) $\varepsilon_2\{2\}$; (d) $\varepsilon_3\{2\}$ as a function of impact parameter in 5.02 TeV Pb+Pb collisions in CGC effective theory. 
  The full lines correspond to the full result, given by Eqs.~(\ref{results1}).
  The dashed line is the approximate value used in~\cite{Giacalone:2019kgg}, where one only keeps the first term in each equation. 
}
\label{Fig:CGC}
\end{figure*}

Figure~\ref{Fig:CGC} displays our result for Pb+Pb collisions at $\sqrt{s_{\rm NN}}=5.02$~TeV.
We have used the value $Q_{s0}=1.06$~GeV, which gives a good fit to LHC data~\cite{Giacalone:2019kgg}. 
The dashed line in each panel represents the contribution of the first term in each line of Eqs.~(\ref{results1}), while the full line is the full result.
As expected from the general discussion following Eq.~(\ref{results1}), the additional terms give a negative correction, which causes a modest reduction of fluctuations for large impact parameters.

\begin{figure}[t]
\centerline{\includegraphics[width=8.cm]{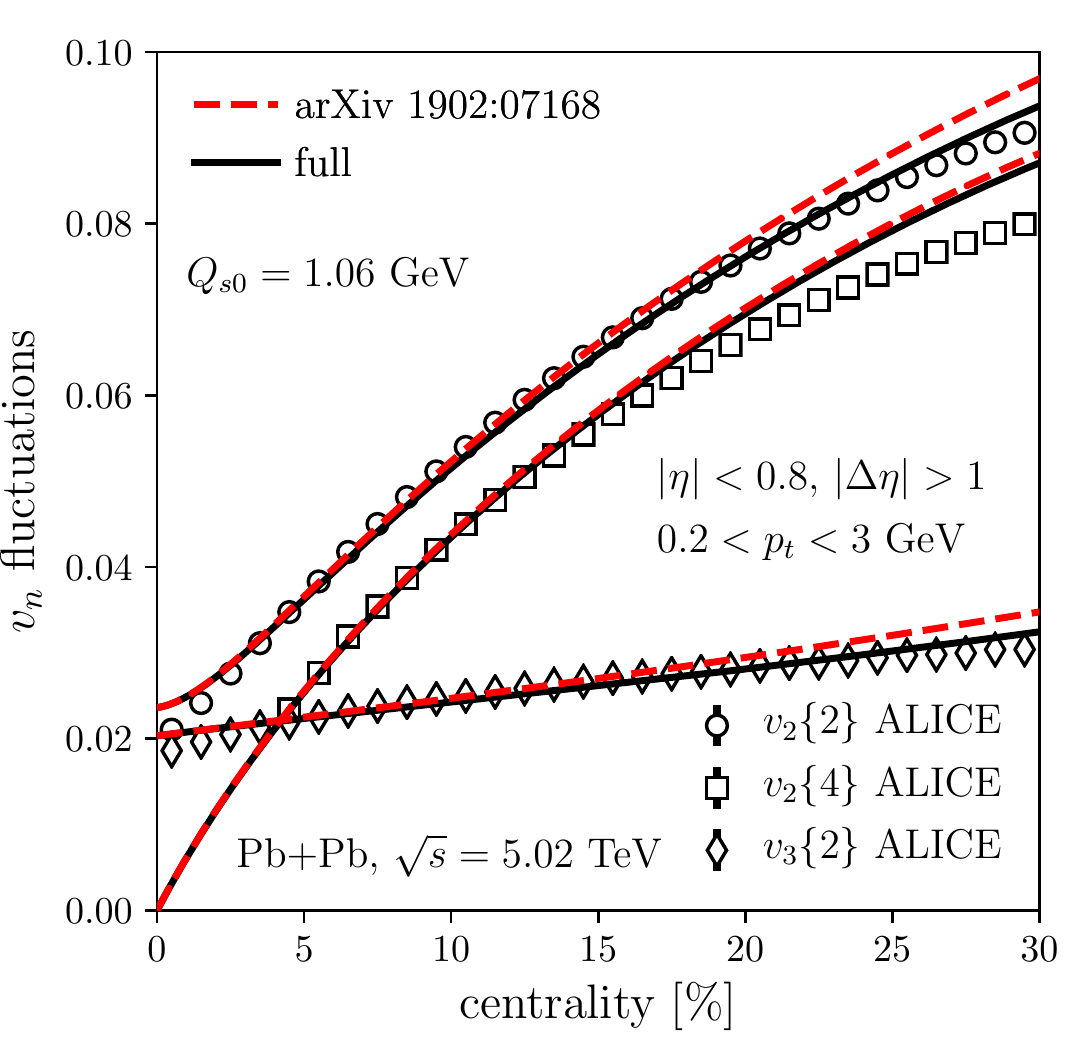}}
\caption{$v_2\{2\}$, $v_2\{4\}$ and $v_3\{2\}$ as a function of centrality. 
  Symbols: ALICE data~\cite{Acharya:2018lmh}. 
  Lines are our CGC calculations of Fig.~\ref{Fig:CGC}, rescaled according to Eq.~(\ref{rescaling}).
  The centrality is defined as $\pi b^2/\sigma_{\rm PbPb}$, with $\sigma_{\rm PbPb}=764$~fm$^2$~\cite{Abelev:2013qoq}.
}
\label{Fig:data}
\end{figure}
Finally, we compare our CGC calculations to experimental data on $v_2\{2\}$, $v_2\{4\}$ and $v_3\{2\}$ in Pb+Pb collisions at ~$\sqrt{s_{\rm NN}}=5.02$~TeV~\cite{Acharya:2018lmh}. 
We use the proportionality relations
\begin{eqnarray}
  \label{rescaling}
v_2\{2\}&=&\kappa_2\varepsilon_2\{2\},\cr
v_2\{4\}&=&\kappa_2\langle\varepsilon_2\rangle,\cr
v_3\{2\}&=&\kappa_3\varepsilon_3\{2\}, 
\end{eqnarray}
where the response coefficients are fitted to data, and 
we have used the property that $v_2\{4\}$ is approximately equal to the mean elliptic flow in the reaction plane.\footnote{The cumulant $\varepsilon_2\{4\}$ actually differs from $\langle\varepsilon_2\rangle$~\cite{Bhalerao:2006tp,Alver:2008zza} but the difference is of the same order as $\bar\varepsilon_2-\langle\varepsilon_2\rangle$, and we neglect it for simplicity.}
We use the same values as in~\cite{Giacalone:2019kgg}, namely, $\kappa_2=0.237$ and $\kappa_3=0.195$.
Results are displayed in Fig.~\ref{Fig:data}.
The subleading terms, which were neglected in~\cite{Giacalone:2019kgg}, improve  the agreement with $v_2\{2\}$ and $v_3\{2\}$ data, but the splitting between $v_2\{2\}$ and $v_2\{4\}$ is too small above 20\% centrality.

Note that these results validate the approximation made in Ref.~\cite{Giacalone:2019kgg}, where only the leading terms of Eqs.(\ref{results1}) were considered.

\section{Modifications at fixed total energy}
\label{s:long}

Working at fixed energy may be a more accurate representation of a tight centrality selection than working at a fixed impact parameter~\cite{Abelev:2013qoq}. 
As explained in the introduction, we assume that the impact parameter is fixed; but we can nevertheless study which modifications occur if the energy is also fixed.

The condition that the total energy is exactly the same for all events does not modify the general analysis carried out in Sec.~\ref{s:pert}, but it does have an effect  on the simplifications made in Sec.~\ref{s:short}.
The reason is that this condition induces a long-range correlation.
If this is the only long-range correlation in the system, it can be accounted for by the following simple modification of the 2-point function~\cite{Blaizot:2014nia,Gronqvist:2016hym}:
\begin{equation}
  \label{econs}
  S({\bf s}_1,{\bf s}_2)
  \rightarrow
  S'({\bf s}_1,{\bf s}_2)\equiv S({\bf s}_1,{\bf s}_2)-\frac{\xi({\bf s}_1)\xi({\bf s}_2)}{\int_{\bf s}\xi({\bf s})},
\end{equation}
where the additional term enforces the sum rule $\int_{{\bf s}}\delta\rho({\bf s})=0$ at the level of the 2-point function:
\begin{equation}
\int_{{\bf s}_1}S'({\bf s}_1,{\bf s}_2)=0.
\end{equation}
Due to the modification (\ref{econs}), Eq.~(\ref{2pointav3}) is modified to
\begin{equation}
  \label{2pointav4}
  \langle\delta f\delta g\rangle=\frac{\int_{\bf s}\xi({\bf s})}
                {\langle E\rangle^2}
\left( \{ fg\}-\{f\}\{g\}\right).
\end{equation}
Note that since the energy is fixed, the integral of $\xi({\bf s})$ no longer represents the variance of energy, as in Eq.~(\ref{defevar}).
Evaluating the contribution of the additional term to Eqs.~(\ref{pertbareps2}), (\ref{pertsigma2}) and (\ref{pertsigma3}), one finds the following modifications to the results (\ref{results1}): 
\begin{eqnarray}
\label{modifE}
  \langle\varepsilon_2\rangle&\to& \langle\varepsilon_2\rangle
+\frac{\int_{\bf s}\xi({\bf s})}
                {\langle E\rangle^2}
                \frac{\{{\bf ss^*}\}
\left(  \{ {\bf s}^2\}\langle {\bf ss^*}\rangle
-\langle {\bf s}^2\rangle \{ {\bf ss^*}\}
\right)}
                     {\langle{\bf ss^*}\rangle^3}\cr
                     \sigma^2&\to&\sigma^2-
\frac{\int_{\bf s}\xi({\bf s})}
                {\langle E\rangle^2}
\frac{\left|  \{ {\bf s}^2\}\langle {\bf ss^*}\rangle
  - \langle {\bf s}^2\rangle\{ {\bf ss^*}\}
  \right|^2}
     {\langle|{\bf s}|^2\rangle^4}\cr
 \varepsilon_3\{2\}^2&\to& \varepsilon_3\{2\}^2.
\end{eqnarray} 
In the specific case of identical, point-like sources, curly brackets and angular brackets coincide and all additional terms vanish. 
This is the reason why Eq.~(\ref{results2}), which is derived without fixing the energy, coincides with the results obtained earlier~\cite{Bhalerao:2006tp,Bhalerao:2011bp} where the total energy was fixed.
In the more general case where $\langle\rho({\bf s})\rangle$ and $\xi({\bf s})$ are different, the additional terms in Eq.~(\ref{modifE}) contribute.
Note that the correction to $\sigma^2$ is always negative, which is intuitive since by fixing the energy, one suppresses one source of fluctuations.

In the case of CGC effective theory, the additional terms in Eq.~(\ref{modifE}) are of much smaller magnitude than the corrections in Eq.~(\ref{results1}).
Specifically, the contributions to $\langle\varepsilon_2\rangle$ and $\sigma^2$ are $8\times 10^{-4}$ and $-3.5\times 10^{-5}$ at $b=10$~fm, respectively, and smaller at smaller impact parameters. 

\section{Conclusions}

We have presented full expressions of the mean and variance of the initial anisotropies $\varepsilon_2$ and $\varepsilon_3$, to leading order in the density fluctuations, as a function of the 1- and 2-point functions of the energy density field $\rho({\bf s})$. 
Each quantity ($\langle\varepsilon_2\rangle$, $\sigma^2$, $\varepsilon_3\{2\}^2$) is dominated by one term. 
Other terms, which were neglected in~\cite{Giacalone:2019kgg}, are only significant for large impact parameters, where they lead to a modest reduction of anisotropic flow fluctuations.
We have also shown that the modifications of these results, depending on whether or not one fixes the total energy of the event, are in practice negligibly small for all centralities. 

\section*{Acknowledgements}
M.L.~acknowledges support from FAPESP projects 2016/24029-6  and 2017/05685-2, and project INCT-FNA Proc.~No.~464898/2014-5.
M.L. and G.G. acknowledge funding from the USP-COFECUB project Uc Ph 160-16 (2015/13).
P.G-R. acknowledges financial support from the `La Caixa' Banking Foundation.
The work of CM was supported in part by the Agence Nationale de la Recherche under the project ANR-16-CE31-0019-02. 
RSB would like to acknowledge the support of the CNRS
LIA (Laboratoire International Associ\'{e}) THEP (Theoretical High
Energy Physics) and the INFRE-HEPNET (IndoFrench Network on High
Energy Physics) of CEFIPRA/IFCPAR (Indo-French Center for the
Promotion of Advanced Research).
RSB also acknowledges the support of the Department of Atomic Energy, India for the award of the Raja Ramanna Fellowship.

\end{document}